\documentclass{ws-mplb}

\begin{document}

\markboth{Lakhno V.D.}
{Phonon interaction of electron in the translation-invariant strong-coupling theory}

\catchline{}{}{}{}{}
%
%%%%%%%%%%%%%%%%%%%%%%%%%%%%%%%%%%%%%%%%%%%%%%%%%%%%%%%%%%%%%%%%%%%%

\title{Phonon interaction of electrons in the translation-invariant strong-coupling theory}
%\footnote{For the
%title, try not to use more than 3 lines.
%Typeset the title in 10~pt Times Roman, uppercase and boldface.}  }

\author{Lakhno V.D.}
%\footnote{Typeset names in
%8~pt Times Roman, uppercase. Use the footnote to indicate the
%present or permanent address of the author.}}

\address{Institute of Mathematical Problems of Biology,\\
Russian Academy of Sciences, Pushchino, Moscow Region, 142290, Russia}
%\footnote{State completely without abbreviations, the
%affiliation and mailing address, including country. Typeset in
%8~pt Times Italic.}\\
%firstauthor\_id@domain\_name\footnote{Typeset author e-mail
%address in single line.} }

%\author{SECOND AUTHOR}

%\address{Group, Laboratory, Address\\
%City, State ZIP/Zone, Country\\
%secondauthor\_id@domain\_name}

\maketitle

%\begin{history}
%\received{Day Month Year}
%\revised{Day Month Year}
%\accepted{(Day Month Year)}
%\comby{(xxxxxxxxxx)}
%\end{history}

\begin{abstract}
A dependence of phonon interaction on the interelectronic distance is found for a translation-invariant (TI) strong-coupling bipolaron. It is shown that the charge induced by the electrons in a TI-bipolaron state is always greater than that in a bipolaron with spontaneously broken symmetry.
\end{abstract}

\keywords{Froehlich Hamiltonian; polaron; correlation length; Coulomb; quark.}

%\section{General Appearance}

%Contributions to {\it International Journal of Modern Physics B}
%will be reproduced by using the author's submitted typeset manuscript.
%It is therefore essential that the manuscript be in its final form, and
%of good appearance. The typeset manuscript should be submitted to the
%publisher in PDF format as well as in its \LaTeX\ format.

%\section{The Main Text}

Considerable attention given to the bipolaron problem in recent years centers around attempts to explain the superconductivity phenomenon with the use of the mechanism of Bose-condensation of bipolaron gas \cite{1}$^,$\cite{2}$^,$\cite{3}. In this context the study of the interaction between the electrons caused by their interaction with phonons is a vital task.  In papers \cite{4}$^-$\cite{6} a new concept of a translation-invariant bipolaron (TI-bipolaron) was introduced which possesses much higher coupling energy, than a bipolaron with spontaneously broken symmetry (SBS-bipolaron). Of interest is to calculate the interaction energy as a function of a distance between the electrons in a TI-bipolaron and to find the value of the charge induced by the electrons in a polar medium.  Notice that for the case of SBS-bipolarons, these points were discussed in a lot of papers \cite{7}$^-$\cite{9}.
Following \cite{4}$^-$\cite{6}, we will proceed from Froehlich Hamiltonian for a bipolaron which in the coordinates of the center of mass has the form:
%(1)
\begin{equation}
\hat{H}=-\frac{\hbar^2}{2M_e}\Delta_R-\frac{\hbar^2}{2\mu_e}\Delta_r+U\left(\left|\vec{r}\right|\right)
+\sum_k\hbar\omega^{ }_ka^+_ka_k+\sum_k2\cos\frac{\vec{k}\vec{r}}{2}\left(V_ke^{i\vec{k}\vec{R}}a_k+H.C.\right)
\label{1}
\end{equation}

\noindent
where $R$, $r$ are coordinates of the center of mass and relative motion of electrons, respectively:
$M_e=2m$, $\mu_e=m/2$, m is an electron mass, $a^+_k$, $a_k$ are operators of a phonon field;
$V_k=(e/k)\sqrt{2\pi\hbar\omega/\tilde{\epsilon}V}$, $\tilde{\epsilon}^{-1}=\epsilon^{-1}_{\infty}-\epsilon^{-1}_{0}$,
 $\omega _k=\omega$ is a phonon frequency, $e$ is an electron charge,
$\epsilon^{-1}_{\infty}$, $\epsilon^{-1}_{0}$ are high-frequency and static dielectric constants,
$V$ is the systems volume, $U(r)=e^2/\epsilon_{\infty}\left|\vec{r}\right|$.

After excluding the center of mass coordinate by means of Heisenberg transformation \cite{9-1}, with the use of Lee-Low-Pines transformation (LLP) \cite{10}:
%(2)
\begin{equation}
S_2=\exp\left\{\sum_kf_k\left(a_k-a^+_k\right)\right\},
\label{2}
\end{equation}

\noindent
the energy of electron-phonon interaction of the electrons $U_{int}(r)$, according to (1), is written as:
%(3)
\begin{equation}
U_{int}(r)=\left\langle 0\left|S^{-1}_2\left(\sum_k2V_k\cos\frac{\vec{k}\vec{r}}{2}
\left(a_k+a^+_k\right)\right)S_2\right|0\right\rangle=
4\sum_kV_kf_k\cos\frac{\vec{k}\vec{r}}{2}.
\label{3}
\end{equation}

\noindent
According to \cite{6}, when the LLP function $f_k$ is chosen in the Gaussian form:
%(4)
$$f_k=-N\vec{V}_k\exp\left(-k^2/2\mu\right),$$
\begin{equation}
\bar{V}_k=2V_k\left\langle \Psi\left|\cos\frac{\vec{k}\vec{r}}{2}\right| \Psi\right\rangle,
\Psi(r)=\left(\frac{2}{\pi l^2}\right)^{3/4}\exp(-r^2/l^2),
\label{4}
\end{equation}

\noindent
where $N$, $\mu$, $l$ are varying parameters, with the use of (3) $U_{int}(r)$ is expressed as:
%(5)
\begin{equation}
\tilde{U}_{int}(\tilde{r})=
-\sqrt{\frac{x^2+16y}{x^2+8y}}\,\frac{1}{\tilde{r}}\,F\left(\frac{2\tilde{r}}{\sqrt{16y+x^2}}\right),
F(x)=\frac{2}{\sqrt{\pi}}\int^x_0e^{-t^2}dt
, \label{5}
\end{equation}

\noindent
where $\tilde{U}_{int}(\tilde{r})=U_{int}(r)/(4me^4/\hbar^2\tilde{\epsilon}^2)$, $\tilde{r}=(e^2m/\hbar^2\tilde{\epsilon})r$
 - are dimensionless variables. The values of $x=x(\eta)$, $y=y(\eta)$
(where $\eta=\epsilon_{\infty}/\epsilon_0$ is the parameter of ion coupling) for each $\eta$ value
are determined from the condition that the function $\Phi(x,y;\eta)$ be minimum \cite{6}:
%(6)
\begin{equation}
\Phi(x,y;\eta)=\frac{6}{x^2}+\frac{20,25}{x^2+16y}-\frac{16\sqrt{x^2+16y}}{\sqrt{\pi}(x^2+8y)}+4\frac{\sqrt{2/\pi}}{x(1-\eta)}
. \label{6}
\end{equation}

\noindent
Relation of quantities $x$, $y$ with the parameters $\mu$, $l$ in (4) is given by formulae  $x=l\alpha$, $y=\alpha^2/\mu$, where:
%(7)
\begin{equation}
\alpha=\left(e^2/\hbar\tilde{\epsilon}\right)\sqrt{m/2\hbar\omega}
, \label{7}
\end{equation}

\noindent
 is a constant of electron-phonon coupling.

\begin{figure}[bt]
%\centerline{\psfig{file=fig1.eps,width=5.65in}}
\centerline{\psfig{file=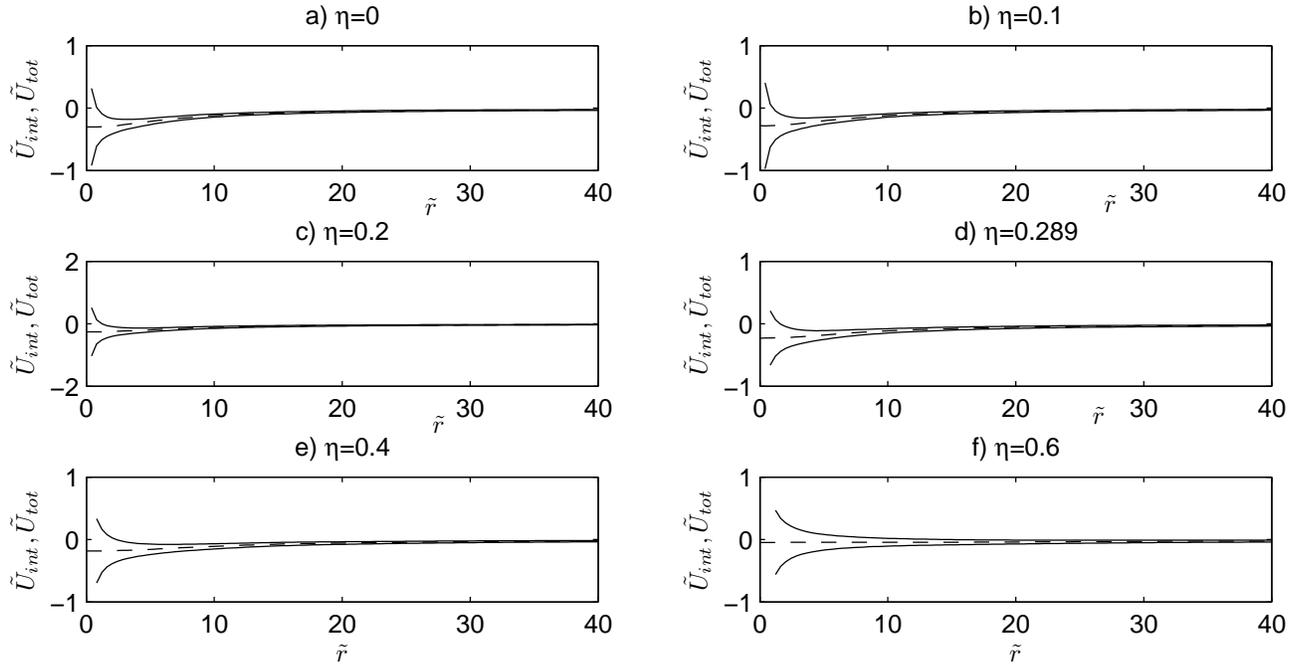,width=8in}}
%\vspace*{8pt}
%\caption{The dependence of interaction potential on $\eta$}
\caption{The dependence of interaction potentials $\tilde{U}_{int}$ (dashed), total potential $\tilde{U}_{tot}$ with positive $U(r)$ in (8), which corresponds to electron Coulomb repulsion (upper solid curve) and negative $U(r)$ (lower solid curve) which corresponds to quark attraction, on $\eta$.}
\end{figure}

Fig.1 demonstrates the dependencies  $\tilde{U}_{int}(\tilde{r})$ for some values of $\eta$ parameter.
It is seen that for small $\tilde{r}$, the interaction potential is independent of $\tilde{r}$,
for intermediator has linear dependence on $r$,
while for large $\tilde{r}$, is has a Coulomb form: $\tilde{U}_{int}(\tilde{r})\sim1/\tilde{r}$.
Fig.1 also suggests that at the point $\eta=\eta_c=0,289$, i.e. at the point where a TI-bipolaron decays into TI-polarons \cite{6}, interaction $U_{int}(r)$ does not demonstrate any jumps and changes continuously as
$\eta$ increases up to the value of $\eta=1-1/2\sqrt{2}$, at which the total energy of a TI-bipolaron
$E_{bp}=\Phi\alpha^2$ vanishes. The total interaction potential $U_{tot}(r)$ should include the Coulomb interaction $U(r)$:
%(8)
\begin{equation}
U_{tot}(r)=U_{int}(r)+U(r)
, \label{8}
\end{equation}
and is shown on Fig. 1 (upper solid curve).

\noindent
It looks like Coulomb interaction in the case of small $r$ and has a near-linear shape in a certain range of $r$ variation (this is especially clear in fig.1 f): $\eta=0,6$). This behavior reminds the interaction between quarks, with repulsive instead attractive as in the case of quark Coulomb potential (Fig. 1, lower solid curve). (A polaron model of quarks was considered in \cite{11}).

The knowledge of $U_{int}(r)$ enables us to calculate the density distribution of a charge $\rho_{ind}(r)$ induced by electrons in a polar medium. Assuming:
%(9)
\begin{equation}
U_{int}(r)=-2e\varphi_{ind}(r)
, \label{9}
\end{equation}

\noindent
where $\varphi_{ind}(r)$ is a potential induced by the electrons, we will write for $\rho_{ind}(r)$:
%(10)
\begin{equation}
\Delta_r\varphi_{ind}(r)=4\pi\rho_{ind}(r). \label{10}
\end{equation}

\noindent
With the use of (5), (9), (10) we express $\rho_{ind}(r)$ as:
%(11)
$$\rho_{ind}(r)=\frac{32}{\pi}\sqrt{\frac{2}{\pi}}
\frac{e}{\epsilon}\left(\frac{me^2}{\hbar^2\tilde{\epsilon}}\right)^3\tilde{\rho}(\tilde{r})$$
\begin{equation}
\tilde{\rho}(\tilde{r)}=\frac{1}{(x^2+16y)\sqrt{x^2+8y}}\exp\left(-8\tilde{r}/(16y+x^2)\right)
. \label{11}
\end{equation}

\noindent
The total charge $Q$ induced by a TI-bipolaron:
%(12)
\begin{equation}
Q=\int\rho_{ind}(r)dV, \label{12}
\end{equation}

\noindent
is equal to:
%(13)
\begin{equation}
Q=\sqrt{\frac{16y+x^2}{8y+x^2}}\;\frac{2e}{\tilde{\epsilon}}. \label{13}
\end{equation}

\noindent
Fig.2 shows the dependence of $f=\sqrt{(16y+x^2)/(8y+x^2)}$ as a function of the parameter $\eta$. Fig.2 suggests that the value of a charge $Q$ induced by the electrons in a TI-bipolaron state is always greater than that of a charge $2e/\tilde{\epsilon}$ induced by the electrons in a SBS-bipolaron state.  These values coincide only for $\eta\rightarrow1-1/2\sqrt{2}$ - value, when the effective distance between the electrons in a TI-bipolaron (correlation length) is equal to infinity. This result suggests that the adiabatic approximation in this limit holds for a TI-polaron too.

\begin{figure}[bt]
\centerline{\psfig{file=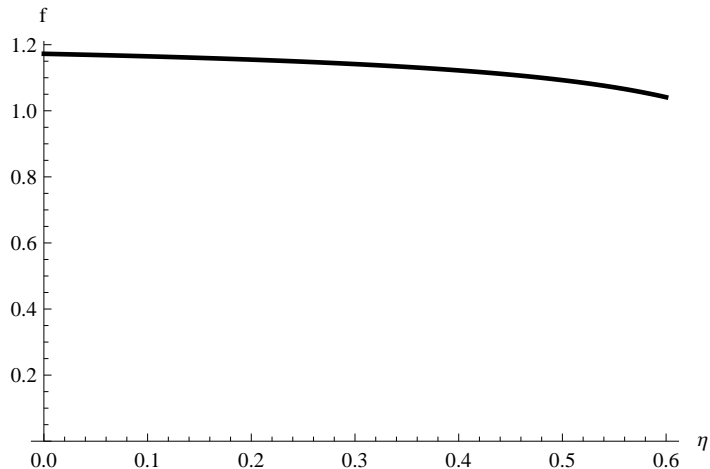,width=3.65in}}
\vspace*{8pt}
\caption{The dependence of function $f=\sqrt{(16y+x^2)/(8y+x^2)}$ on $\eta$}
\end{figure}

In conclusion it may be said that the frequently introduced concept of interpolaronic interaction \cite{2}$^, $\cite{7}$^-$\cite{9} in the case of a TI-bipolaron is objectless, since even for large $r$, a TI-bipolaron cannot be presented as the one consisting of two individual polarons (here an analogy with confinement of quarks is appropriate).

This presentation, however, can be sensible, if for a certain value ($\eta=\eta_c$) a decay of a TI-bipolaron into two individual TI-polarons is possible. For this case, the dependence of the polarons interaction on the distance was calculated in \cite{9}.

\section*{Acknowledgements}

%This section should come before the References and Appendices.
%Funding information may also be included here.
The work was supported by RFBR, project N 13-07-00256.

\end{document}